\documentclass[letterpaper]{article} 
\usepackage{aaai2026}  
\usepackage{times}  
\usepackage{helvet}  
\usepackage{courier}  
\usepackage[hyphens]{url}  
\usepackage{tikz}
\usetikzlibrary{positioning}
\usepackage{graphicx} 
\usepackage{natbib}  
\usepackage{caption} 
\usepackage{algorithm}
\usepackage{algorithmic}
\usepackage{amsmath}
\usepackage{booktabs}
\usepackage{newfloat}
\usepackage{listings}
\DeclareCaptionStyle{ruled}{labelfont=normalfont,labelsep=colon,strut=off} 
\floatstyle{ruled}
\newfloat{listing}{tb}{lst}{}
\floatname{listing}{Listing}
\title{Chatsparent: An Interactive System for Detecting and Mitigating Cognitive Fatigue in LLMs}

\author {
    Riju Marwah\textsuperscript{\rm 1,2},
    Vishal Pallagani\textsuperscript{\rm 2},
    Ritvik Garimella\textsuperscript{\rm 2},
    Amit Sheth\textsuperscript{\rm 2}
}
\affiliations {
    \textsuperscript{\rm 1}Guru Gobind Singh Indraprastha University, India\\
    \textsuperscript{\rm 2}Artificial Intelligence Institute, University of South Carolina, USA\\
}

\begin{document}

\maketitle

\begin{abstract}

LLMs are increasingly being deployed as chatbots, but today’s interfaces offer little to no friction: users interact through seamless conversations that conceal when the model is drifting, hallucinating or failing. This lack of transparency fosters blind trust, even as models produce unstable or repetitive outputs. We introduce an interactive demo that surfaces and mitigates \textit{cognitive fatigue}, a failure mode where LLMs gradually lose coherence during auto-regressive generation. Our system, Chatsparent, instruments real-time, token-level signals of fatigue, including attention-to-prompt decay, embedding drift, and entropy collapse, and visualizes them as a unified fatigue index. When fatigue thresholds are crossed, the interface allows users to activate lightweight interventions such as attention resets, entropy-regularized decoding, and self-reflection checkpoints. The demo streams live text and fatigue signals, allowing users to observe when fatigue arises, how it affects output quality, and how interventions restore stability. By turning passive chatbot interaction into an interactive diagnostic experience, our system empowers users to better understand LLM behavior while improving reliability at inference time. The demo video is available at \url{https://youtu.be/ktqkZyYWDDE}.

\end{abstract}


\section{Introduction and Contributions}

Rapid deployment of LLMs in conversational interfaces has made them appear natural, seamless, and effortless \cite{lebeuf2017software}. Yet, this very lack of friction hides a fundamental risk: users are encouraged to place blind trust in outputs, even when the model is drifting, hallucinating, or failing \cite{georgiou2025chatgpt}. Current chatbot interfaces provide little transparency into when such degradation occurs, leaving users unaware of why answers become repetitive, incoherent, or overconfident.

We argue that these behaviors are not rare anomalies but symptoms of a deeper failure mode \cite{arbuzov2025beyond}, which we call \textit{ cognitive fatigue}, a gradual loss of coherence and stability as autoregressive generation unfolds. Fatigue emerges naturally from architectural pressures such as decaying prompt attention \cite{li2024measuring}, accumulation of hidden-state drift \cite{wei2025shadows}, and entropy collapse in the token distribution. Crucially, fatigue can be detected online during inference and mitigated without retraining, turning auto-regressive generation from a passive risk into an active control problem.

To address this, we introduce an interactive demo called Chatsparent that makes cognitive fatigue visible, measurable, and actionable. Our system instruments three lightweight, token-level signals (attention-to-prompt decay, embedding drift, and entropy collapse) and fuses them into a unified \textit{fatigue index}. This index is streamed in real time alongside model outputs, giving users a live view of when fatigue arises. When thresholds are crossed, the interface enables \textit{retrain-free interventions} including prompt reinsertion, entropy-regularized decoding, and self-reflection checkpoints. By letting users toggle decoding and interventions, the demo turns chat into a diagnostic that exposes model dynamics and improves reliability. Our contributions are threefold: \textbf{(a)} we formalize and measure \textit{cognitive fatigue} as an online state using token-level signals of attention decay, embedding drift, and entropy collapse; \textbf{(b)} we introduce retrain-free interventions that counteract fatigue during decoding; and \textbf{(c)} we provide an interactive demo system that visualizes fatigue in real time and allows users to apply these interventions, transforming generation into a more reliable and interpretable process.

\begin{figure}[h!]
    \centering
    \includegraphics[width=0.48\textwidth]{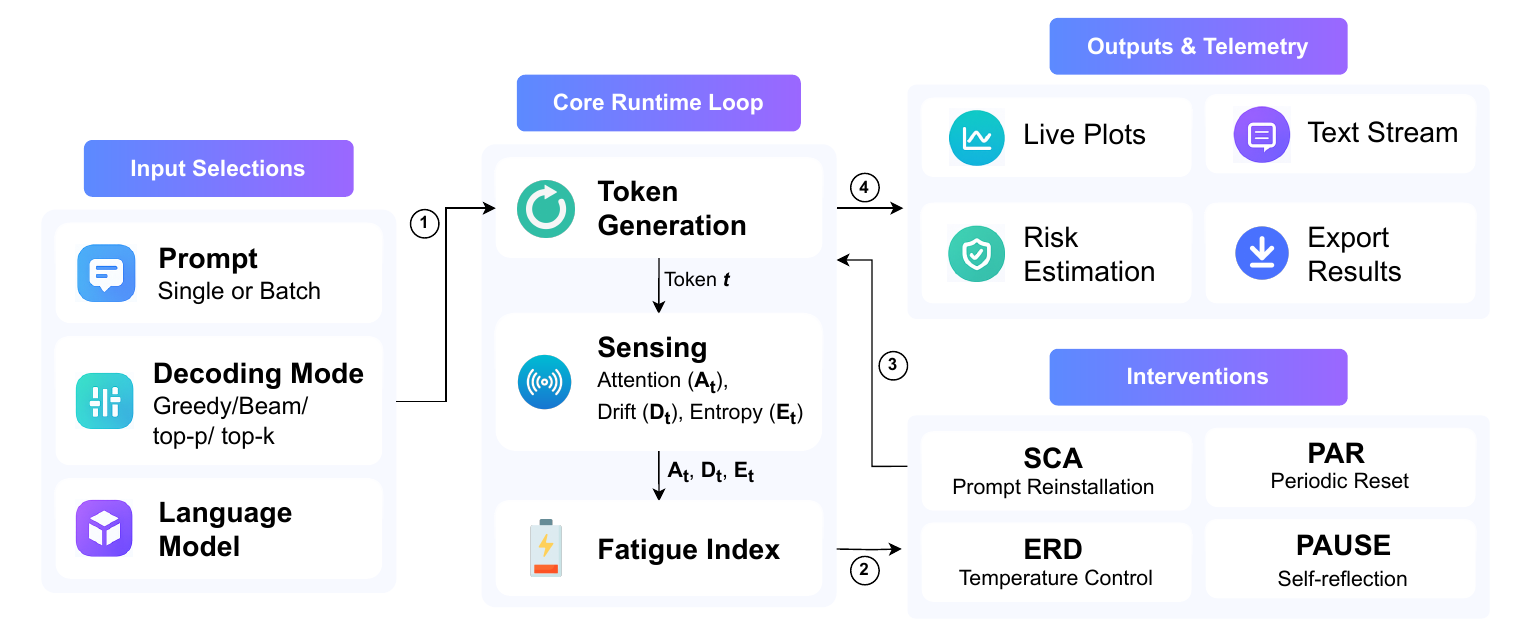}
    \caption{Fatigue-aware decoding pipeline.}
    \label{fig:example}
\end{figure}

\section{System Overview}

We view autoregressive decoding as a controlled process with a latent reliability state. At each step $t$, we compute three signals: $A_t$ is the mean last-layer attention mass from the current token to a fixed prompt slice; $D_t = \lVert h_t - h_0 \rVert_2$ measures embedding drift from the prompt’s last-token hidden state $h_0$; and $E_t$ is the entropy of the next-token softmax. After normalizing each via $\phi_{\cdot} \in [0,1]$, we define a compact fatigue index
\[
F_t = w_A \, \phi_A(A_t) \;+\; w_D \, \phi_D(D_t) \;+\; w_E \, \phi_E(E_t),
\]
with a small hysteresis band to avoid rapid toggling. Weights are defaulted as $w_A = 0.40,\; w_E = 0.35,\; w_D = 0.25$ 
(attention prioritized for instruction-following; entropy for calibration; drift down-weighted due to cross-model scale). A simple policy monitors $F_t$ (or any constituent signal) and triggers an intervention when the safe region is breached. The demo surfaces this control loop as a single pipeline consisting of three stages: Sense (signals), Decide (thresholds and hysteresis), and Intervene (SCA, PAR, ERD, PAUSE).

\section{Signals and Interventions}
\textbf{Signals.} Attention-to-prompt captures loss of instruction focus as context grows; a sustained decline indicates prompt under-use. Embedding drift reflects the cumulative effect of attention/MLP updates in the shared residual stream; rising values indicate wandering representations. Entropy tracks calibration: persistent lows reveal over-confidence and a tendency toward repetition. Together, the signals form a lightweight proxy for the onset of fatigue.

\noindent \textbf{SCA (attention-triggered prompt reinsertion).} When the attention signal $A_t$ falls below a threshold and a cooldown permits, we re-prepend the original prompt and keep only a short recent tail of tokens so the sequence stays within the context limit. This “break-glass” action refocuses the model without editing the key–value cache \cite{rawte2024sorry}. 

\noindent \textbf{PAR (periodic attention reset).} At a fixed cadence $k$, the context is rebuilt as $[\text{prompt} + \text{recent\_tail}]$. PAR produces bumps in attention around reset boundaries, acting as a preventive nudge against gradual decay. \cite{xu2024lookback}. 

\noindent \textbf{ERD (entropy-regularized decoding).} At each step, we measure $E_t$ and adjust the temperature $T \in [T_{\min}, T_{\max}]$ to track a target entropy $H_{\text{target}}$: if the entropy is too low, increase $T$; if too high, decrease it. ERD curbs entropy collapse and indirectly flattens attention decay while leaving the representation dynamics largely unchanged. \cite{liu2023lost}. 

\noindent \textbf{PAUSE  (Self-reflection checkpoints, chain-of-thought questioning)} On a fixed cadence \textit{r} or whenever uncertainty or drift increases (for example, when the entropy signal $E_t$ falls outside its safe band or when the drift signal $D_t$ rises above a threshold), the model briefly pauses generation to perform a targeted self-check. \cite{ezzeldin2024positional}. 

\section{Experimental Snapshot}
We evaluate \texttt{Falcon-7B-Instruct} \cite{almazrouei2023falcon} quantized to 4-bit NF4 (\texttt{bitsandbytes}), with eager attention. Decoding defaults: top-$p=0.95$, $T=1.0$, $\texttt{max\_new}=120$. We probe every token to log three online signals: attention-to-prompt $A_t$, embedding drift $D_t$, and next-token entropy $E_t$. These are combined into the Fatigue Index with an entropy ``healthy band'' $[1.5, 3.0]$ and light hysteresis for stability. The interventions are as follows:
\begin{itemize}
  \item SCA: $\tau_A=0.010$, cooldown $=8$, max $=1$, $\texttt{tail\_keep}=128$.
  \item PAR: $\texttt{reset\_every}=50$, $\texttt{tail\_keep}=128$.
  \item ERD: fix top-$p$; adjust $T \in [0.7, 1.5]$ with gain $k=0.35$ to track $H^* = 2.8$.
  \item PAUSE: insert a 1-line focus check every 30 tokens (gate $=5$ tokens).
\end{itemize}

\noindent We use HotpotQA (dev) \cite{yang2018hotpotqa}. For the demo table we report a representative single item under the free-form prompt; a strict short-answer variant is provided in the supplement. Metrics shown are Mean FI (averaged over generated tokens) and wall-clock latency.

\begin{table}[h!]
\centering
\begin{tabular}{lccc}
\toprule
\textbf{Method} & \textbf{Mean Fatigue Index $\downarrow$} & \textbf{Latency (s) $\downarrow$} \\
\midrule
Baseline & 0.36 & 213.47 \\
ERD      & 0.31 (-0.05) & 212.45 \\
PAR      & 0.34 (-0.02) & 222.36 \\
PAUSE      & 0.31 (-0.05) & 228.02 \\
SCA    & 0.32 (-0.04) & 225.11 \\
\bottomrule
\end{tabular}
\caption{Comparison of fatigue index and decoding latency for a single-prompt setting.}
\end{table}

\section{Demo Walkthrough}
From the left control panel, paste a prompt or pick a HotpotQA item, choose decoding (Greedy/Top-k/Top-p/Beam) and model, then optionally enable SCA/PAR/ERD/PAUSE and set their knobs. The center panel streams the answer, shows a Fatigue Index gauge, plots the three signals and explains the run; you can overlay the baseline and export CSV/JSON. The right panel reports risk of degradation. 

\begin{figure}[h!]
    \centering
    \includegraphics[width=0.48\textwidth]{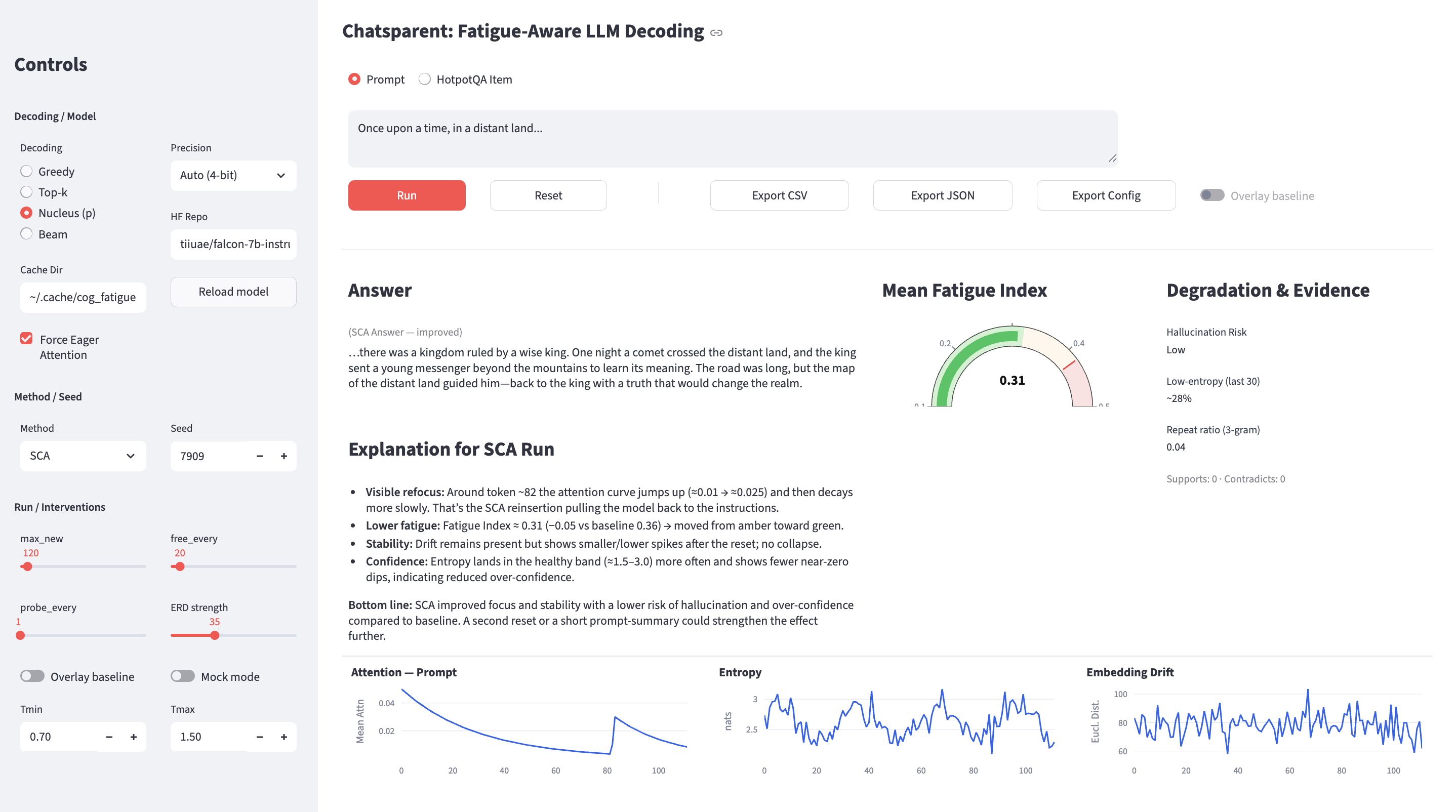}
    \caption{Interactive demo.}
    \label{fig:example}
\end{figure}


\section{Conclusion}
We operationalize cognitive fatigue as a detect-and-act loop during decoding. By monitoring attention decay, representation drift, and entropy collapse, and by applying simple interventions, we improve long-horizon reliability on multi-hop QA without retraining. The demo offers a practical path toward safer, longer-context deployments by turning silent degradation into visible signals and actionable control.

\newpage
\bibliography{aaai2026}

\end{document}